\documentclass[9pt,twocolumn,twoside]{osajnl}

\journal{ol} 

\setboolean{shortarticle}{true}
\usepackage{graphicx,amssymb,mathtools,mathptmx} 

\title{    Exploring the ultimate limits:  Super-resolution enhanced by partial coherence     }

\author[1*]{Zden\v{e}k~Hradil}
\author[1] {Dominik ~Koutn\'y  }
\author[1]{Jaroslav~\v{R}eh\'a\v{c}ek}

\affil[1]{Department of Optics, Palack\'y University, 17. listopadu 12,
  771 46 Olomouc, Czech Republic}

\affil[*]{Corresponding author: hradil@optics.upol.cz}

 \dates{Compiled \today}  

\ociscodes{(270.0270) Quantum optics;
  (270.5585) Quantum information and processing;
  (120.3940) Metrology.} 

\begin{document}

\begin{abstract}
 The resolution of separation of two elementary signals  forming a partially coherent superposition, defined  by quantum Fisher information  and normalised with respect to detection probabilities, is always limited by the  resolution of incoherent mixtures.  
    However, when  the  partially coherent superpositions   are prepared in a controlled way  the precision can be enhanced by up to   several orders of magnitude above this limit.   Coherence also allows the sorting of  information about various parameters into distinct channels as demonstrated  by parameter of  separation  linked with the  anti-phase superposition and  the centroid position linked with the  in-phase superposition. 
\end{abstract} 

\setboolean{displaycopyright}{true}
\maketitle

\section{Introduction}


The quest for ultimate limits serves as  unfathomable source  of  inspiration for theoretical as well as experimental research.  
The "ultimate limits"  usually  do not represent a  final goal   but rather show the direction  worth pursuing. It is intriguing to note that coherence (and coherent states)  provides one of the  keystones  for  quantum optics but the role of (partial) coherence in metrology   has not been fully exploited yet.

 Recent developments have been   inspired by reconsideration of well-known  concepts of classical optics such as Rayleigh   \cite{Rayleigh} or Sparrow  \cite{Sparrow} resolution criterion from the point of view of quantum estimation theory  \cite{Tsang-review} and particularly the  concept of  Fisher information \cite{Helstrom}.   The problem of classical resolution can be paraphrased as the question  of how well can we distinguish two
bright spots, or more generally two  elementary signals, in which is coded the information about the parameter of interest.
The celebrated classical resolution criteria  suggest that this can be done up to the
distances when two blurred spots start to overlap.
{ This rule of thumb can be justified by the  analysis of   the intensity pattern, since the Fisher information vanishes quadratically for  separations approaching zero. }
As shown by Tsang and coworkers \cite{Tsang} and demonstrated experimentally on different platforms 
\cite{Steinberg,Lvovsky,Tang,Opt, TF}, this behaviour can be avoided if quantum estimation theory is
adopted for the estimation of geometrical parameters, namely the transversal
separation and the centroid positions of two equally bright spots with known 
intensities.
In this context, the Fisher information refers to quantum measurements and
converts to the Quantum Fisher Information (QFI) upon optimising over all
possible measurement schemes  \cite{Helstrom, Liu20}.  
The role of coherence in quantum estimation problems has been discussed  in  recent  publications   \cite{Saleh,Comment_Tsang,Reply_Larson} without leading to a consensus.
  More light was shed on this problem in the paper \cite{coherence} showing that if  the separation between two sources is coded into the coherent superposition state, the QFI may become unlimited.  Such an effect, however,  is always associated with  the measurement on the dark channel with vanishing signal.     In this Letter  we extend   the theory to partially coherent states addressing several issues.
At  first  we show that there are non-equivalent models  for the description of partially coherent states built on the  mixing of normalised states or  alternatively on the  mixing of coherent amplitudes, when  any fusion of  both those generic models is allowed as well.  When the 
 strength of the signal is taken   into account,  none of the models presents a metrological advantage since precision of  any of those partial superpositions is limited  by the precision of incoherent mixtures though the  Quantum Fisher Information itself may become infinite.  
 {            In this sense our  conclusion can be seen as an optical illustration of the profound and more general observation that  probabilistic metrology or any post-selection  scheme can never improve quantum limits on estimation of a single parameter, though some estimates may  appear  occasionally excellent  \cite{Caves2014}.   In addition to this }
  we show, how the limiting  precision of incoherent mixtures can {be overcome} by special  state preparation.
 This  will be  achieved by careful tailoring of the degree of partial coherence depending on the separation. This scenario  surely lacks practical relevance for estimating an unknown separation,    {  but this is a common issue  for all  model-dependent   schemes,  when the structure cannot be assumed ad hoc as unknown without additional verification steps.   The example  with enhanced resolution is  just stressing  the importance of the "classical" part  of the signal- namely its strength- which  may depend on the estimated parameter. This can be the  result of some  (presumed) internal process responsible for generation of such a signal, or  a part of the  controlled experiment:  In the first state-preparation step, parameters are coded into the signal and prior information about parameters is used.   In the second detection step, where we use only knowledge of the model without using knowledge of the parameters, these parameters are  retrieved from data with an increased precision.
   Such a protocol can  be viewed as  testbed for experimentalists,  and  requires the mastering of several advanced steps   including state engineering,  detection on demand,  noise control and data processing. 
  In this perspective, the coherence spreads information into different interfering channels, into the (modulated) strength or probability distribution.     As a part of this scenario    we will  finally show  that coherence may allow efficient   sorting of information about parameter of  separation  into  the  anti-phase superposition,  and  centroid position  into the   in-phase superposition. 
The analysis   will  clarify  some    conceptual misunderstandings  related to use of unnormalised states    in  quantum information processing \cite{Saleh,Comment_Tsang,Reply_Larson}.  }   
 In the following,  the quantum formalism of Dirac notation will be adopted, though the motivation for this research stems from classical optics.  The method used here will rely on quantum optimisation  and was developed in the context of quantum theory. The conclusions derived here   will be still valid in the broader context of quantum metrology.   To highlight the optical interpretation, the  relevant conclusions will be framed in terms of classical optics.

\section{Method and Results}

 Let us  review some basic facts from  quantum estimation theory for unbiased estimators of a variable $\theta.$ The  Fisher Information $F$ quantifies the content of
information per detected particle; the Cr\'amer--Rao inequality   (CR) \cite{Helstrom} (here for a
single parameter $\theta$  and sequential detection),
\begin{equation}
  (\Delta \theta )^2\geq  1/  {\cal H } \equiv \frac{1}{{n  {\cal F}}}\,,
\label{precision}
\end{equation}
sets a bound given by the precision ${\cal H} $ with which $\theta$ can be estimated from
the data.  Here, $(\Delta \theta )^2$ is the expected value of the variance of the unbiased
estimator, and $n$ is the number of repetitions of  the detection. { The Fisher information $ \cal F$  plays the pivotal role in quantum metrology and can be calculated for the given model by means symmetric logarithmic derivative (for more information see the topical review \cite{Liu20},  where the mathematically rigorous formulation is illustrated with numerous examples  for applications in quantum science). }  The Fisher information quantifies  well the   resources needed for reaching some target precision: small values of  $ \cal F$  must be compensated by an  increase of the data set.
As argued in  \cite{coherence}   special  attention must be paid to coherent superpositions, where some extra cost are associated with preparation of a particular superposition of displaced states:   
\begin{eqnarray}
  \label{Phi12}
 & |\Phi_{1,2}\rangle&=\frac{1}{2}\bigl(|\Psi_{+}\rangle
                                  \pm e^{i\varphi}|\Psi_{-}\rangle\bigr)  ,  
                                  |\Psi_{\pm}\rangle =\exp(\pm iPs/2)|\Psi\rangle,
\end{eqnarray}        
 where $P$ is  the  operator of transverse momentum, $s$ parametrises the separation, {   $\varphi = 0$  and 1D geometry is assumed for the simplicity.  The superposition state  with $\pm$ sign  will be called in-phase and anti-phase   $ |\Phi_{1,2}\rangle, $respectively.}
 As reviewed in the Supplementary Materials, costs for preparing  such  states      are simply given by  the corresponding norms $  | \Phi_{1,2} |^2. $   
 The normalisation   used here $    | \Phi_{1} |^2 +   | \Phi_{2} |^2 = 1  $  guarantees that costs  $C $  for preparing the incoherent mixture equal to one.     In general, the costs  for preparing the  partial coherent mixture  enter the  CR inequality  as   a multiplicative rescaling $n$ into $nC$  in the  inequality  Eq. \ref{precision} with the obvious meaning:  if $n$  pairs of the separated  states  $|\Psi_{\pm}\rangle $ are at our disposal, the desired superposition   is generated $nC $ times.   Mathematical  models  leading to specific choice of  $C$ will be specified later. 
 


 
Partially coherent states  and  the corresponding rate factors  can be generated by  many ways, all of them   legitimate in the sense of estimation theory but  just a few of them useful.  The  following  two basic  models  (A, B)    are  useful for understanding of the role of  coherence in quantum metrology.   In  Model   A, the  states are normalised before they are added, whereas in Model B  the unnormalised states ( "amplitudes")    are added. The way how to sort out those states from a generic scheme is clarified in Supplementary Materials  \cite{Supp}. 
        \begin{eqnarray}
\label{state_A}
   \rho_A =    \frac{p_1}{||  \Phi_1 ||^2}     |\Phi_1\rangle \langle \Phi_1  |    + \frac{ p_2}{||  \Phi_2 ||^2}     |\Phi_2\rangle \langle \Phi_2  |  ,  \\
 \label{state_B}  
  \rho_B =  \frac{1}{C}  \biggl[ {p_1}   |\Phi_1\rangle \langle \Phi_1  |    +   p_2     |\Phi_2\rangle \langle \Phi_2  |  \biggr] ,  \\
  \label{rate} 
  C = p_1  ||  \Phi_1||^2   +  p_2  ||  \Phi_2||^2 , \quad p_1 + p_2 = 1 .
   \end{eqnarray}
  The rate  $C$ is equal in both the cases.
  As shown in Supplementary Materials, such states can be generated "blindly," meaning  without the knowledge about the true value of the  estimated parameter of separation $s.$  { The difference between these  models  was a subject of some misunderstanding. }
    The theory elaborated by  Larson and Saleh  in Refs. \cite{Saleh,Reply_Larson}  represents a mixture of models A and B, whereas Tsang and Nair  \cite{Tsang-review,Comment_Tsang}  have  considered just  Model B as the only option relevant for optics.   We also point the reader to the arguments  from  the Appendix 3 of the Ref. \cite{Tsang-review}  based on Helstrom book \cite{Helstrom} advocating the use of unnormalised  function of mutual coherence  instead of its normalised form.  This may be  but need not be always so, depending whether the normalisation factor $C$ is or is not available from the measurement. {  These issues are detailed  in Supplementary Materials \cite{Supp}.  As  a brief  message for experimentalists, it  matters how the partially coherent state  is created - either by mixture of  coherent  superpositions or  by an incoherent mixture superimposed with coherent signal, whether or not the intensities are fixed or depend on parameters, and  particularly the   strength of the detected  signal is important.  Coherent effects allow to distribute information in different  channels (i.e. partially coherent states) and/or modulation of the strength of the signal.    "Rayleigh's curse" reappears  only if some piece of this information is omitted but coherence cannot be blamed for that. }


{   The central  role  in quantum metrology is played by  QFI matrix  associated with the given model. Though this is in general  a rather involved problem, the off-the-shelf formulae are at disposal   \cite{Liu20}.   For the understanding of the role of coherence, the  rank-2 density matrix  is enough and QFI can be cast in the form }
   \begin{eqnarray}
 {\cal F}_{2-rank} =  \sum_{i=1,2} \frac{ (\partial _s \lambda_i)^{2}}{\lambda_i}  + 4  \sum_{i=1,2}   \lambda_i  [ \frac{ ||\partial \Phi_i  ||^2 }{  ||\Phi_i ||^2 } - \frac{ |\langle  \Phi_i | \partial \Phi_i \rangle |^2     }{ ||\Phi_i ||^4}  ] .
\label{QFI-2}
\end{eqnarray}
where $\lambda_{1,2}$ are  the eigenvalues of the density matrix  and    $|\partial \Phi\rangle  $  is the derivative of the state with respect to the parameter. As an  important distinction the 
  QFI  in  Model A is given just by  adding  Fisher informations of  individual coherent superpositions, 
whereas   in  Model B  an extra term  appears as a   consequence of modulating  eigenvalues of those coherent superpositions. { This  contribution is large for states close to  anti-phase superposition,  where the signal is vanishing for small separation,   see Supplementary Materials  \cite{Supp}. } However,  if the QFI is accompanied  by the   rate factor C  (\ref{rate}) ,  the overall precision is limited by the resolution of the incoherent mixtures.
 
This limit can be overcome in some specific scenarios when the parameter  is coded into the signal in order to distinguish it more efficiently afterwards.  {  Assume  the scheme when the partially coherent state is created in such a way  that  normalised  superposition states    $|\Phi_1\rangle , |\Phi_2 \rangle $  are normalised and  mixed  with  the parameter -dependent weights  $\bar{ p}_1(s),  \bar{ p}_2(s).$ If we choose 
\begin{eqnarray}
\bar{ p}_1(s)  = \frac{ p_1  ||\Phi_1||^2 }{   C  }, \quad  \bar{ p}_2(s) = \frac{ p_2  ||\Phi_2 ||^2 }{  C  } , \\
  \rho_E =    \frac{ \bar{p}_1}{||  \Phi_1 ||^2}     |\Phi_1\rangle \langle \Phi_1  |    + \frac{\bar  {p}_2}{||  \Phi_2 ||^2}     |\Phi_2\rangle \langle \Phi_2  |     = \rho_B,  \\
  C = p_1  ||  \Phi_1||^2   +  p_2  ||  \Phi_2||^2 , \quad     \bar{p}_1 + \bar{p}_2  = p_1 + p_2 = 1 ,
  \end{eqnarray}
  we get the same state as in  Model  B! What has changed is just  the rate factor  which will be given as 
     \begin{eqnarray}
   \label{E-rate} 
  C_E = \frac{1}{C} [  p_1  ||  \Phi_1||^4   +  p_2  ||  \Phi_2||^4 ] ,
   \end{eqnarray}
 with  all the coefficients having the same meaning in   all  models.    QFI  remains large  even when multiplied by  the rate factor  $C_E.$  Here the  product $ C_E  {\cal F}$ scales  with the small  separation like $1/s^2 $ and may exceed considerably  the precision of an incoherent mixture.     This is the essence of models with enhanced resolution  (Model E).  This behaviour is exemplified in    Fig. 1  in comparison to  Model B, which has just indicative meaning  and demonstrates  that such resolution is limited by incoherent superposition.   }
  
  Though this result might appear paradoxical, there is even some  classical  rationale behind it  as explained by behaviour of    ``anti-phase''
superpositions,  which may  be considered as always "distinguishable"  due to the gap between the peaks \cite{Cesini},
but such a signal is vanishing when approaching anti-phase. 
{ If we analyse the resolution in Model E in dependance on the  weight parameter  $ p , p_1 = p, p_2= 1-p $  (see Supplementary Materials for details) \cite{Supp}, we conclude that maximum  is achieved for states  close to the mixtures when  in-phase component   $   \Phi_1$ is taken with the weight of anti-phase phase component, i. e  $ | \Phi_2|^2   $ and the other way around.  This defines  "nearly optimal" point on the plot in Fig. 1 close to maximum of resolution.   For small separations,  the anti-phase component is vanishing and in-phase component  is dominating, hence the optimal state requires to combine  strong signal with vanishing weight and vanishing signal with strong weight factors. This explains the  "equal roles"   of both the  "anti-phase"  and "in-phase" components  in models yielding enhanced resolution.}
 For  such a state  $C_E = 1/2,$  what means that  just half of the signal in comparison to incoherent mixture is used. However,   the increase of QFI  is enormous and  significantly outperforms the precision  of QFI for incoherent mixtures.

 \begin{figure}
  \includegraphics[width=1\columnwidth]{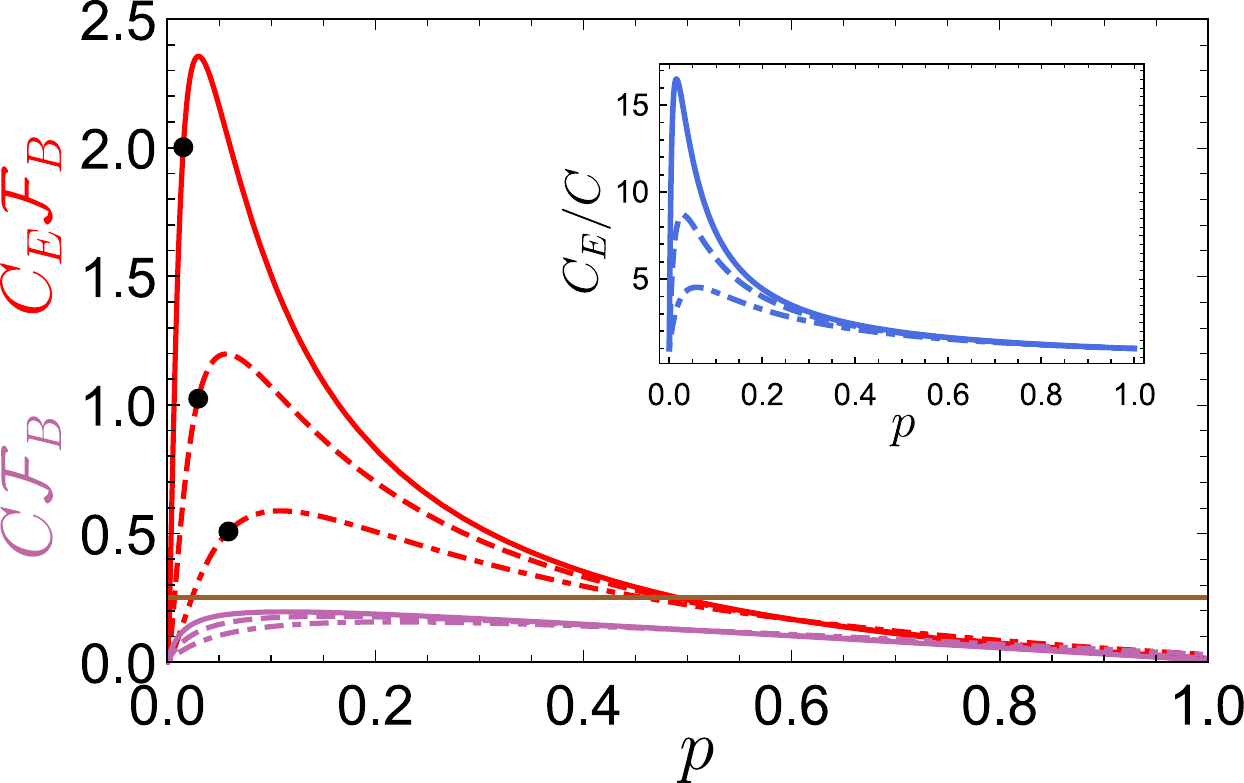}
  \caption{
     Dependence of the  precision   $C {\cal F}  $   on  the weights $   p_1= p, p_2=1- p$    of partially coherent mixtures for   values of the separations  $ s =  0.5, 0.7, 1  $   {   (dashed-doted, dashed, full ) listed from the bottom up.}  Red curves  indicate enhanced resolution (E).  { The purple curves  corresponding to  Model B  are  below  the threshold  $1/4$-   the limit  of incoherent mixtures, and have just indicative meaning. }  Black points  correspond to the "near optimal" $ p$ -values, where $C = 1/2, $ see the explanation in the text.  The displacement is in units of $\sigma, {\cal  F}$ is in units of
    $\sigma^{-2}$  and numerical calculations are done for Gaussian distribution. The inset shows the ratio of  corresponding rate factors C  {  for separations  $ s =  0.5, 0.7, 1  $  listed from the bottom up. }    }
\end{figure}

 To exemplify the enhancement in numbers let  us consider the  parameters   at the separation   $ s= 0.1 \sigma $   where  super-resolution has been already demonstrated   on different   experimental platforms \cite{Opt,TF} for incoherent mixtures.
 For optimal enhanced super-resolution  one can get the numerical factors $C= 0.02 ,    C_E =0.67,  $ whereas   $ {\cal F} = 89 $ and the precision scales  with the factor $   C_E   {\cal F} \approx  59.$  In comparison   the precision  for the incoherent mixtures of separated sources  is constant    $  (\Delta P)^2 = 0.25 .$ 
  More detailed calculations are provided in the Supplementary Materials.
 
    The  enhanced  resolution need not be the only  potential benefit of coherence manifested by constructive and destructive interference.  Coherent effects  may also facilitate new ways for sorting of information.      We  demonstrate how to exploit the  full information content   of the signal  including the information contained in the modulated  rate factor  $C$- the norm of the state,  and in the normalised state.
      As calculated in Supplementary Material \cite{Supp}   the full information carried by unnormalised state  $ | \Phi \rangle  $  {    consists of  QFI for  a pure  normalised state  (e.g. given by Eq. ( \ref{QFI-2} )  for single eigenvalue  $ \lambda_1 = 1 $ ),  rescaled by the  normalisation   $  C =  || \Phi ||^2 $   with  an extra added Fisher information  term corresponding  to the  modulation  $  {(\partial C)^2}/{C},$   all together   }
      \begin{eqnarray}
{\cal  F}_{total}  = 4 ||\partial \Phi ||^2   + \frac{    [   \langle  \Phi | \partial \Phi \rangle -    \langle  \partial \Phi| \Phi \rangle        ]^2     }{ ||\Phi ||^2} .
\end{eqnarray}
   Applying this result to specific in-and anti-phase superpositions $|\Phi_{1,2}\rangle $  we  conclude  that   for small separations  the Fisher information for the  separation is carried dominantly by  the anti-phase component, and particularly is coded into the norm,  whereas the information about centroid position is carried  by the  in-phase superposition. This simple observation  may serve as inspiration  for future complex protocols for distributing and sorting of information by means of multi port devices.    { This effect is graphically illustrated in  Fig.2  plotting the Fisher information for separation of Gaussian profiles for     in-phase superposition $ {\cal F}_1$ and   anti-phase superposition  ${\cal F}_2, $   and particularly, its   dominant part  corresponding to  the Fisher information  attributed to the norm of anti-phase component   $  {\cal F}_{C}.$} 

 {
 \begin{figure}
  \includegraphics[width=1\columnwidth]{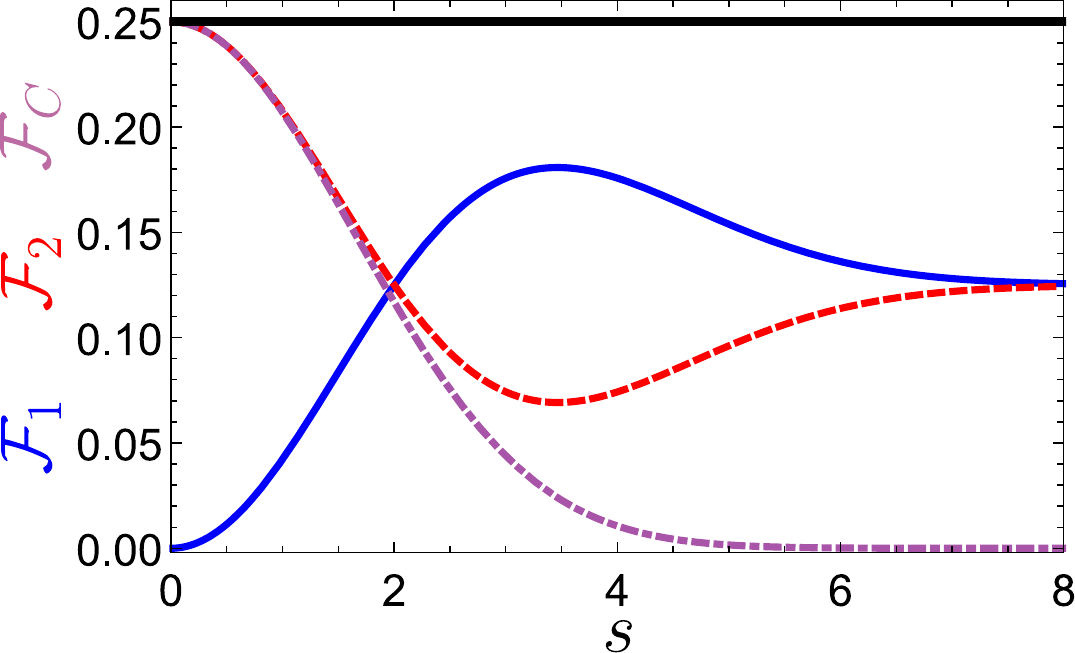}
  \caption{
  Fisher information for separation as a function of separation for   in-phase superposition $ {\cal F}_1$  (full blue)   and  anti-phase  superposition  $ {\cal F}_2$  (dashed red).    Dominant contribution of $  {\cal F}_2 $  for small separations stems from the modulation of  the norm  of anti-phase superposition   plotted as $ {\cal F}_C$  (dashed-dotted purple).   The  sum  of individual terms    $ {\cal F}_1$ and $ {\cal F}_2$  is  conserved and saturates the  limiting  resolution for  incoherent mixture given as $1/4.$  The separation  is in units of $\sigma,  $  ${\cal F}$ is in units of  $\sigma^{-2}$ and numerical calculations are done for Gaussian distribution. }
\end{figure}        
          }

\section{Discussion}

 The essence of the proposed scheme  for enhanced resolution hinges upon the  manifestation of  genuine  partial coherence,  {  when    both the  in- and anti-phase  coherent channels  and their intensities  depend on the  estimated parameter.    The enhancement of the precision stems from the strong  modulation of the intensity terms  near the point  of the almost vanishing signal.  The scenario with particular choice of weights  considered here serves as an example of how the modulation of coherent signals can be utilised for enhancing the precision  above the limit of incoherent mixtures.  
 There  is even   a certain classical   rationale  behind  this effect.   As shown in Refs. \cite{tempering, reading}  points where the intensity of the  PSF is vanishing  contribute to   increased precision  even in the  case of   the intensity detection. The improvement is   manifested by the linear scaling of Fisher information.  The theoretical analysis  presented here is actually the quantum analogue of the same  effect:  since the signal must vanish for any observation, the signal itself must  vanish.   Such arguments cannot be used for estimating the separation between unknown sources; however 
 the analysis is relevant for  coding the information into the prepared state  which will serve for reading out the information afterwards. }

Though the effect of enhanced super-resolution is theoretically possible, it will be challenging to observe it experimentally.  Several  steps must be mastered  simultaneously:  precise  state engineering of  phase  sensitive states,  phase sensitive measurements as projection on the eigenstates of the symmetric logarithmic derivative,  and all  this must be done with  extremely  weak fields. These steps  have been implemented  to a certain extent separately on various experimental set-ups.   In particular, the experiments in  Refs. \cite{tempering,reading} demonstrated that it is possible to detect  faithfully the zero intensity signal  and  based on this  to see the region of linear scaling of classical Fisher information.  New progress of sophisticated optical methods using  quantum pulse gates  \cite{QPG} or mode multiplexing \cite{multiplexer} are promising tools for  implementation of the phase sensitive detection on demand.


 
 \section*{Conclusions}
 
  The analysis of partial coherence identifies    metrological advantages of future   protocols inspired by quantum information processing.   The ultimate precision of incoherent mixtures provides a natural bound which  can  be overcome in controlled experiments with  information  additionally coded  into the partially coherent signal.  Coherence may also   provides additional benefits in the form of sorting of information for  different parameters into different channels as demonstrated on   the  separation attributed to  the anti-phase component  and the  centroid position  linked with the in-phase component.

\noindent 
{\bf Acknowledgments}
The authors are indebted to Hubert de Guise for valuable comments.
We acknowledge
financial support from the Grant Agency of the Czech Republic (Grant
No.~18-04291S), the Palack\'y University (Grant
No. IGA\_PrF\_2020\_004)   the project ApresSF  supported be the MEYS, Czech Republic,  under the QuantERA programme, which has received funding from the European Union's Horizon 2020 research and innovation programme " and    H2020-FETOPEN-2018-2019-2020-01 StormyTune.

\noindent 
{\bf Disclosures}  The authors declare no conflicts of interest.

\noindent 
See Supplement  for supporting content.

\newcommand{\DOI}[2]{\href{#1}{#2}}

\newpage

\section{Supplementary material}

 	Coherent superpositions   \cite{coherence} play an essential role for   understanding of partial coherence.
 	In order to assess the cost needed  for preparation of such  states assume in  the following  the state of our system entangled with  "spin-like" ancilla states 
 	\begin{eqnarray}
 	\label{entangled}
 	&  |\varphi\rangle&=2^{-1/2}\bigl(|\Psi_+\rangle
 	\otimes|\mathnormal{\uparrow}_x\rangle
 	+e^{i\varphi}|\Psi_-\rangle\otimes|\mathnormal{\downarrow}_x\rangle\bigr)\\
 	&&= |\Phi_1\rangle\otimes|\mathnormal{\uparrow}_z\rangle
 	+|\Phi_2\rangle\otimes|\mathnormal{\downarrow}_z\rangle\,,\\
 	& |\Phi_{1,2}\rangle&=\frac{1}{2}\bigl(|\Psi_{+}\rangle
 	\pm e^{i\varphi}|\Psi_{-}\rangle\bigr),\\
 	 &\quad|\Psi_{\pm}\rangle &=\exp(\pm iPs/2)|\Psi\rangle.
 	\end{eqnarray}

 	As argued in \cite{coherence}   the state $  |\Phi_{1,2}\rangle$  is generated by projecting the ancilla onto the basis  $|\mathnormal{\uparrow}_z\rangle, |\mathnormal{\downarrow}_z\rangle .$    This happens with the probability   rate $   C  =  ||\Phi_{i}||^2 \equiv \langle \Phi_i | \Phi_i \rangle , i= 1,2   .$       Similarly, the marginal distribution over the ancilla  selects the state of the system in the incoherent mixture. Due to the normalisation used here this happens with the  rate  equal to one establishing   a reference for later comparison.   If  the rate is included  as the multiplicative factor into the  CR inequality,  all  results are normalised with respect to resources - or equivalently with respect to detection probabilities-rather than conditioned by detected events. The differences may become  significant in the cases when e.g. QFI is  diverging but the detected channel is actually dark. 
 	The special choice  of coherent superposition states  with   $\varphi = 0 $ is of particular interest.  In both cases, the  QFI for  the in-phase as well as anti-phase superpositions 
 	scales quadratically to zero.  In addition to this, the  rate for the  anti-phase component vanishes as well and  it might thought that such a basis is not suitable for super-resolution. As will be seen, the opposite is true.

 	There are two simple models  on how to create partially coherent states that can be   constructed  directly from coherent superpositions.  Model  A  is defined as incoherent  superposition of projectors into the basis of  coherent superposition  states     
 	\begin{eqnarray}
 	\rho_{A} =  \frac{p_1}{||  \Phi_1 ||^2}     |\Phi_1\rangle \langle \Phi_1  |    +   \frac{p_2}{||  \Phi_2 ||^2}     |\Phi_2\rangle \langle \Phi_2  | , \\
 	C = p_1  ||  \Phi_1||^2   +  p_2  ||  \Phi_2||^2 .
 	\label{rate_SM}
 	\end{eqnarray}
 	This sorting scheme   requires mixing  the  results of the projections  into the ancilla states $ |\mathnormal{\uparrow}_z\rangle\, |\mathnormal{\downarrow}_z\rangle\ $  with prior probabilities  $ p_1, p_2.$  
 	Yet  another mixed superposition of signal state can be projected   from the entangled state  in Model B,   if the ancilla is conditioned by  the detection   
 	\begin{eqnarray}
 	\Pi_B =  p_1 |\mathnormal{\uparrow}_{z}\rangle \langle \mathnormal{\uparrow}_{z}| +  p_2  |\mathnormal{\downarrow}_{z}\rangle \langle \mathnormal{\downarrow}_{z}|, 
 	\end{eqnarray}
 	yielding the state  
 	\begin{eqnarray}
 	\label{state_B}
 	\rho_B=   \frac{1}{    p_1||  \Phi_1 ||^2 +   p_2 ||  \Phi_2 ||^2 }    [    p_1 |\Phi_1\rangle \langle \Phi_1  |    +  p_2  |\Phi_2 \rangle \langle \Phi_2  |    ], 
 	\end{eqnarray}
 	with the same rate factor  Eq. (\ref{rate_SM}) as  in Model A .
 	The  subtle distinction between variants A and B is of fundamental importance.  It explains some  confusion  in the discussions    \cite{Comment_Tsang,Reply_Larson}  and  provides arguments for the enhancement of the super-resolution.  Let us note in passing that    Model  A  assumes  the summation of the normalised states (as  done in quantum optics), wheres in  Model B   "amplitudes"  (like in coherent optics) are added.
 	
 	Two specific models  have been  considered recently for   construction of partially coherent states. 
 	Following the arguments in  Larson and Saleh   \cite{Saleh} , the  partially coherent state  can be cast  as the superposition of fully coherent  (here $\varphi$ dependance can be  included)  and fully incoherent parts
 	\begin{equation}
 	\rho_{LS} =  \frac{p}{ ||\Phi_1||^2} |\Phi_1\rangle \langle \Phi_1 | +   \frac{1-p}{2}\{  |\Psi_+\rangle \langle \Psi_+ | + |\Psi_-\rangle \langle \Psi_- |  \}.
 	\end{equation}
 	Due to  the identity $$    |\Phi_1\rangle \langle \Phi_1 | + |\Phi_2\rangle \langle \Phi_2 |= \frac12[  |\Psi_+\rangle \langle \Psi_+ | + |\Psi_-\rangle \langle \Psi_- | ]$$
 	the state can cast in the  form 
 	\begin{eqnarray}
 	\rho_{LS} =  \bigl[ \frac{p}{ ||\Phi_1||^2 }     + 1-p \bigr] |\Phi_1\rangle \langle \Phi_1 | +  (1-p)  |\Phi_2\rangle \langle \Phi_2 | .
 	\end{eqnarray}
 	Such a state corresponds to  the mixture between the generic  Models A and B. 
 	
 	On the other hand,  Tsang  and Nair  \cite{Comment_Tsang}  were motivated  by an   optical definition  considering the partially mixed state  with the degree of coherence $\gamma$ 
 	{
 		\begin{equation}
 		\rho_{TN} =  N_0  \bigl[     |\Psi_+\rangle \langle \Psi_+ | +   |\Psi_-\rangle \langle \Psi_-|    +  \gamma    |\Psi_+\rangle \langle \Psi_- |     +  \gamma^*    |\Psi_- \rangle \langle \Psi_+ |     \bigr].
 		\end{equation}
 		Using the  polar representation for the  degree of coherence  $\gamma = |\gamma |  e^{ - i \varphi} $ and  the definition of  in- and anti- phase superposition states,}  the  partially mixed state can be  simply rewritten to the form  of  Model  B 
 	\begin{eqnarray}
 	\rho_{TN} =  N_0   \bigl[    (2 - |\gamma|)     |\Phi_1\rangle \langle \Phi_1 | +   2 (1-|\gamma| )  |\Phi_2 \rangle \langle \Phi_2 |    \bigr]  , \\
 	N_0^{-1} =   2 - |\gamma|  - |\gamma|  || \Phi_2 ||^2 .
 	\end{eqnarray}
 	All models   {  including their mixtures } are legitimate,  but as they  correspond to different preparations,  they cannot be confused.

 	{ The    normalisation  of the states represents another issue in recent discussions.  As argued  by Tsang in Ref. \cite{Comment_Tsang},  the  unnormalised  function of mutual coherence should be used  instead of its normalised form. 
 		We  agree that  this argument can be justified  in some cases  but the correct usage of QFI, however, requires to use  normalised states in order to  guarantee the consistent  normalisation of probability distributions. 
 		Particularly  in  Model B  the factor $C$  represents   just the  trace of unnormalised density matrix  in the state Eq. (\ref{state_B}).  If this factor is  known as a function of estimated parameters,  it  can be exploited for estimation and  the total Fisher information carried by such a signal reads 
 		\begin{eqnarray}
 		{\cal  F}_{total}  =  \frac{(\partial_s C)^2}{ C}  + C {\cal F}_n.
 		\label{total}
 		\end{eqnarray}
 		The first term corresponds to (classical) Fisher information  attributed to $C$  and ${\cal F}_n $ represents  QFI  of the  (normalised) quantum  state.  Here and in the following we will  use the term Quantum Fisher Information just for  normalised states.  Before resorting to calculations of QFI for 2-component system let us evaluate the total Fisher information for the pair of in-and anti- phase superpositions $|\Phi \rangle_{1,2}. $ As derived in 
 		\cite{coherence}, the QFI for the normalised state reads (for $i=1,2$)
 		\begin{equation}
 		{\cal F }_i = \frac{ 4 ||\partial \Phi_i  ||^2 }{  ||\Phi_i ||^2 } - \frac{ 4  |\langle  \Phi_i | \partial \Phi_i \rangle |^2     }{ ||\Phi_i ||^4} , \quad 
 		C= \langle \Phi_i | \Phi_i \rangle ,
 		\label{QFI_n}
 		\end{equation}
 		and  therefore the total  precision  (\ref{total}) scales with  the  Fisher information  
 		\begin{eqnarray}
 		{\cal  F}_{total}  = 4 ||\partial \Phi_i  ||^2   + \frac{    [   \langle  \Phi_i | \partial \Phi_i \rangle -    \langle  \partial \Phi_i | \Phi_i \rangle        ]^2     }{ ||\Phi_i ||^2} .
 		\end{eqnarray}
 		Adopting usual condition $\langle  P   \rangle  = 0, $   { the Fisher information  for separation associated with the  in-  and anti-phase  superpositions can be easily found as (notice the interchange of indices here)
 			\begin{eqnarray}
 			{\cal F}_1 =   \langle \Phi_2  |(\Delta P) ^2 |\Phi_2 \rangle, \quad 
 			{\cal F}_2 =   \langle \Phi_1  |(\Delta P) ^2 |\Phi_1 \rangle,  \\
 			{\cal F}_1  + {\cal F}_2  = \langle \Psi  | (\Delta P) ^2  | \Psi \rangle  .
 			\end{eqnarray}
 			We note that  for small separations  the anti-phase component  $\Phi_2   $ carries the dominant information about separation, whereas information  embedded in the in-phase component is negligible.  However, sum of both  contributions saturates  the limit of  incoherent mixtures.   In addition to this,   the dominant contribution to $ {\cal F}_2 $ stems from the normalisation  term 
 			\begin{eqnarray}
 			{\cal F}_C =    \frac{(\partial_s C)^2}{ C} .
 			\end{eqnarray}
 			This behaviour is illustrated in Fig. 2 of the main text. Notice that  the  Fisher information for anti-phase superposition  is constant even for vanishing signal  if  a Poissonian noise model is assumed.  
 			Similar argumentation  with complementary results can be applied to the  estimation of the  centroid position  $c_0$ induced by the overal  unitary transformation  $e^{-ic_0 P} .$ The Fisher information terms   for individual channels  are  given as  
 			\begin{eqnarray}
 			{\cal F}_{1}^{cent} =   4  \langle \Phi_1  |(\Delta P) ^2 |\Phi_1 \rangle, \quad 
 			{\cal F}_{2}^{ cent} =   4  \langle \Phi_2  |(\Delta P) ^2 |\Phi_2 \rangle .
 			\end{eqnarray}
 			Obviously,  the dominant contribution is carried by the  in-phase  superposition $\Phi_1.$  }
 		It is intriguing to note that  a beam splitter creating both  superpositions 
 		acts like the device sorting  the components with (dominant)  information dedicated to different parameters.  Such  a distribution  of information   may  provide  another potential advantage to metrology inspired by the quantum information protocols. }

 	{   The analysis of partial coherent fields  requires to use  the concept  of  quantum  Fisher information matrix and symmetric logarithmic derivative  $\cal L$
 		\begin{equation}    
 		\partial \rho = \frac12 [   \rho  {\cal L} +   {\cal L}   \rho ] .
 		\end{equation}
 		This is not an easy task in  the case of  generic  multi-parameter estimation.  The theoretical background,  existing calculation techniques  and   applications in physics has been addressed in the topical review \cite{Liu20}.  The analysis  provided here simplifies to  the case of  rank-two states (see  Eq. (14)  of Ref.  \cite{Liu20})
 		evaluated in the  diagonal basis of 
 		two-component mixed state
 		\begin{equation}
 		\rho = \lambda_1 |u_1\rangle \langle u_1 |  +  \lambda_2 |u_2\rangle \langle u_2 | .
 		\end{equation}
 	}   QFI   can be  written in the form of decomposition into  two  operator  -  ${\cal S}, $  acting in the 2-dimensional space spanned by  the eigenbasis,   and its  orthogonal   part ${\cal P}$ 
 	\begin{eqnarray}
 	{ \cal L}_{\theta}   =   {\cal S}+ {   \cal P } \\
 	{\cal  F}   = Tr(\rho {\cal L}_{\theta}^2) \\
 	{\cal S}  = \sum _k  \frac{ \partial_s \lambda_k}{\lambda_k} |u_k\rangle \langle u_k |  + 2 (\lambda_1-\lambda_2) [\langle u_2| \partial u_1 \rangle   |u_2 \rangle  \langle u_1 |  + h.c. ]
 	\nonumber \\
 	{ \cal P } =  2 [   |u_1^{\perp }  \rangle \langle u_1 | +   |u_2^{\perp }  \rangle \langle u_2 | +  h.c.    ], 
 	\nonumber\\
 	|u_i^{\perp}  \rangle =   | \partial u_i \rangle  - ( |u_1\rangle \langle u_1 | + |u_2\rangle \langle u_2 |)  | \partial u_i \rangle. 
 	\nonumber
 	\end{eqnarray}
 	Here $ | \partial u_1^{\perp} \rangle $ denotes  the projection of the derivative of the eigenstate into the subspace orthogonal to the eigenbasis $ |u_i \rangle .$  
 	The optimal measurement is given by  projections into the eigen-basis of logarithmic   derivative; this will not be discussed here as we simply adress  ultimate limits.  
 	Finally QFI can be brought to the form 
 	\begin{eqnarray}
 	&{\cal  F}_{2-rank}&=  \sum_i \frac{(\partial_s\lambda_i)^{2}}{\lambda_i} + 4 (\lambda_1 -\lambda_2)^2\times\\
 	&& \times | \langle \partial u_1 | u_2 \rangle |^2  
 	+   4 \lambda_1 || \partial u_1^{\perp}||^2  +  4   \lambda_2 || \partial u_2^{\perp}||^2 ,\\
 	&|| u_i^{\perp}||^2 &=  || \partial u_i ||^2 - | \langle u_1| \partial u_i \rangle |^2 -   | \langle u_2| \partial u_i \rangle |^2  .
 	\end{eqnarray}
 	In the following we will specialize to  the common choice  $ \varphi = 0 $ and such  states  for which  $ \langle \psi |e^{is P} | \psi \rangle  $  is real ( including  e.g. Gaussian PSF).  This  assumption simplifies the analysis considerably  since  the states are orthogonal  $  \langle \Phi_2|\Phi_{1}  \rangle  =  \frac{-i}{2} \Im \langle e^{is P} \rangle  =0$
 	providing the eigenstate basis 
 	\begin{eqnarray}
 	|u_{1,2} \rangle =   \frac{1} {\sqrt{    ||\Phi_{1,2} ||^2  }} |\Phi_{1,2} \rangle. 
 	\end{eqnarray} 
 	QFI can be  simplified to the final form 
 	\begin{eqnarray}
 	{\cal F}_{2-rank} =  \sum_{i=1,2} \frac{(\partial_s\lambda_i)^{2}}{\lambda_i}  + 4  \sum_{i=1,2}   \lambda_i  [ \frac{ ||\partial \Phi_i  ||^2 }{  ||\Phi_i ||^2 } - \frac{ |\langle  \Phi_i | \partial \Phi_i \rangle |^2     }{ ||\Phi_i ||^4}  ] .
 	\end{eqnarray}
 	The terms can be easily interpreted. The first term is attributed to the dependance coded in the eigenvalues, whereas the second sum represents the  linear combination of the  QFI for  coherent  superpositions $   |\Phi_{1,2} \rangle $ 
 	given by expression (\ref{QFI_n}).
 	
 	If the eigenvalues are independent of the estimated parameter as in  Model A, the first sum is zero and QFI is just the sum of particular contributions saturating the convexity  
 	condition for QFI. Such a partially coherent state does not offer any advantage for enhancing the precision.
 	
 	An extra term appears in   Model B  as a consequence of  the modulation of eigenvalues $\lambda_{1,2} , $    and this term  can be  large  for small separations
 	\begin{eqnarray}
 	{\cal F}_{\lambda}  =  \sum_{i}  {(\partial_s \lambda_i)^{2}}/{\lambda_i} 
 	=   \frac{ [ \Im \langle  Pe^{isP}  \rangle]^2  } { 1- c^2}     \frac{ p(1-p)}{(||\Phi_2||^2 + cp)^2}.\\
 	C =   ||\Phi_2||^2 + c p,  
 	\end{eqnarray}
 	The parameters  here are denoted for brevity  as    $ p_1 = p, p_2 = 1-p,  c =  \Re \langle e^{isP} \rangle, \quad ||\Phi_2||^2  = \frac12 (1-c) \approx  \frac14 s^2  (\Delta P)^2. $
 	However,  the term $  {\cal F}_{\lambda} $ is  large  but when normalised  with respect to the rate factor  C.   This will be limited by the resolution of incoherent mixture $ (\Delta P)^2.$ 
 	This is an expected result since any detection on the  entangled state Eq. \ref{entangled} is limited by its QFI,  which equals to   $ (\Delta P)^2$  \cite{coherence}. Moreover  the complementary measurement  to  $\Pi_B$
 	\begin{eqnarray}
 	\Pi_{\bar B} = p_2 |\mathnormal{\uparrow}_{z}\rangle \langle \mathnormal{\uparrow}_{z}| +  p_1  |\mathnormal{\downarrow}_{z}\rangle \langle \mathnormal{\downarrow}_{z}|, 
 	\end{eqnarray}
 	sorts the complementary state $\rho_ {\bar B}$ and the complementary rate $C_{\bar B}$ obtained by simple interchange $p_1 \leftrightarrow p_2.$
 	It is intriguing to note that combining those  normalised  states with their rate factors  (what is effectively the combination of un-normalised states) gives again the incoherent mixture of separated states, and henceforth 
 	\begin{eqnarray}
 	C_B  {\cal F}_B  + C_{\bar B}  {\cal F}_{\bar B } \le (\Delta P)^2. 
 	\end{eqnarray}
 	These arguments shows  that (partial) coherence by itself does not bring any advantage with respect to precision when normalised to detection probabilities.
 	For single channel the optimal regime   approaching the resolution of incoherent mixture requires  the adjustment  of the weight $p$ for the given separation $s.$    As can be shown by simple analysis the maximum of the product $C  {\cal F}_{\lambda} $ this  is approximately achieved for $p \approx  \sqrt {\frac{ ||\Phi_2||^2} {c}}  $  with the rate  $  C = \sqrt {{||\Phi_2||^2} {c}} .$ For instance, for  $ s= 0.1 (0.5) \sigma $ the rate factor  C shows that just 2 \% ( 12\%) of the intensity   of incoherent signal  is enough for reaching the limiting precision!

 	This is   the clue to enhance  the resolution  as established in  Model E. What is only   needed is to increase the rate factor C for the state as in    Model B.  This can be done if the state of  Model B  is prepared according to the recipe of   Model A!  In other  words, such a  state can be prepared as the mixture with  weights dependent on the separation 
 	\begin{eqnarray}
 	\bar {p} _i =  \frac{ p_i  ||\Phi_i||^2 }{   C  }.
 	\end{eqnarray}
 	This will give  effectively the same state as in  Model B but with the enhanced rate 
 	\begin{eqnarray}
 	\label{E-rate_SM} 
 	C_E = \frac{1}{C} [  p_1  ||  \Phi_1||^4   +  p_2  ||  \Phi_2||^4 ] .
 	\end{eqnarray}
 	Straightforward analysis  shows the enhancement of precision for 
 	$p, s   \rightarrow 0. $  QFI normalised with respect to the strength of the signal scales  as
 	\begin{eqnarray}
 	{\cal F}_{\lambda}  C_E  \propto  \frac{1}{s^2}.
 	\label{main}
 	\end{eqnarray}
 	The procedure of Model E shows  how to enhance  the information about the estimated parameter   high above the level of incoherent mixtures.  
 	However such a state preparation is not "blind" in the sense that  prior knowledge about the estimated parameter is required and used in the state preparation stage in controlled experiments.

\end{document}